\documentclass[aps,prd,twocolumn,showpacs,preprintnumbers,amsmath,amssymb]{revtex4-1}

\usepackage{graphicx}
\usepackage{dcolumn}
\usepackage{bm}
\usepackage{natbib}
\setcitestyle{square}

\usepackage{amsmath}
\usepackage{braket}
\usepackage{mathrsfs}
\usepackage{amsfonts}
\usepackage{color}

\newcommand{\angstrom}{\mbox{\normalfont\AA}}

\begin{document}


\title{Effect of Ionized Impurity Screening on Spin Decoherence at Low and Intermediate Temperatures in GaAs}

\author{Gionni Marchetti}
\email{marchetti@fhi-berlin.mpg.de}
\author{ Irene D'Amico}
\email{irene.damico@york.ac.uk}
\affiliation{%
Department of Physics, University of York, Heslington, York, YO10 5DD, UK\\
}%




\date{\today}

\begin{abstract}

We study the effect of charged impurity screening on spin decoherence in bulk {\it n}-type GaAs, and analyse in details the effect of the use of different Born approximations applied to a linearized Thomas-Fermi screening theory. The spin relaxation times are calculated by ensemble Monte Carlo techniques, including electron-electron, electron-impurities, and electron-phonons scattering. We carefully choose a parameter region so that all the physical approximations hold, and, in particular, a Yukawa-type potential can be used to describe the screened Coulomb interaction and the Born series converges. Our results show that including the second order Born approximation yields much shorter spin relaxation times compared to the commonly implemented first Born approximation: spin relaxation times may be reduced by hundreds of picoseconds, with the first Born approximation overestimating results by 30\% or more for a large region of parameters. Though our ensemble Monte Carlo simulations include electron-electron and electron-phonon interactions, when considering low to intermediate carrier densities and $T > 50 $ $\mathrm{K}$, but $T$ smaller than the Fermi temperature, our results are in good agreement with Dyakonov-Perel theory when this includes electron-impurity interactions only, which supports this to be the most relevant scattering mechanism for bulk GaAs in this low-intermediate temperature regime.
\end{abstract}

\maketitle

\section{Introduction} \label{intro}

In bulk semiconductors, despite the intricacy of  many-body interactions,  the static screened Coulomb interaction  between a carrier and a shallow impurity, assumed as a point-like particle, is usually described by a Yukawa-type potential \cite{chattopadhyay1981}. The validity of this two-body short-range potential strongly relies on the random phase approximation (RPA) and also on the Thomas-Fermi approximation, which is indeed a static limit for small momentum wavevectors of the dielectric function \cite{ashcroft1976}. These are well-known results of
the interacting Fermi gas theory \cite{giuliani2005}.

The range and strength of the screened Coulomb interaction is then  determined by the screening length, that can be
calculated within a finite temperature linearized Thomas-Fermi approximation (LTFA) in different Born approximations \cite{sanborn1995}.
At room or higher temperatures the inverse screening length in first Born approximation (B1) is a good enough approximation which provides reliable Coulomb
scattering rates, and therefore accurate computation of semiconductor properties such as electron mobility \cite{jacoboni1989} and  spin relaxation
time (SRT) \cite{marchetti2014,marchetti2014a}.
We recall here that B1  is a high-energy approximation \cite{omnes1963}  and for bulk semiconductors  at low/intermediate
temperatures it may become invalid as found, for example, by  Meyer and Bartoli in their study of
electron mobility in {\it n}-type  GaAs \cite{meyer1982} and Silicon \cite{meyer1981} at $T \approx 5 \div 80$ $\mathrm{K}$. However if the Born series converges, and B1 approximation fails, including the second Born
approximation (B2) is usually rewarding because it gives a greater computational accuracy and also sometimes it may reveal new physical insights which were
hidden by a poor approximation \cite{omnes1963}.

A very good approximation for the Born series up to the second term is provided by Schwinger variational principle for
the scattering amplitude \cite{joachain1987}. In the case of Born series formulation for scattering phase shifts this principle,
together with Friedel sum rule (FSR) \cite{friedel1958,stern1967}  gives an analytical expression of the inverse screening length in B2 approximation  in
the limit of low temperatures \cite{patterson1989}. Because the screening determines the strength of Coulomb scattering, its accurate estimate is of
paramount importance for material properties  which strongly depend on Coulomb collisions. This is indeed the case of electron mobility at low
temperatures or at high doping concentration \cite{chattopadhyay1981} or  spin relaxation of an electron ensemble due to spin-orbit coupling in semiconductors
lacking inversion symmetry, as groups III-V and II-VI \cite{zutic2004}. This spin relaxation process in such semiconductors is the Dyakonov-Perel (DP)
mechanism \cite{dyakonov1971}.

 Electron spin decoherence in solid-state systems is  a central theme in spintronics whose main goal is the active manipulation
of spin degrees of freedom for various potential applications: spin-based qubits for quantum information
processes \cite{loss1998}, spin-based devices such a spin field-effect transistor \cite{datta1990}, magnetic tunnel junctions \cite{tsymbal2003}, devices enriched with new functionalities from spin phenomena such as the spin-hall effect \cite{jungwirth2012}, etc. In this regard, GaAs-based semiconductors have been the object of an extensive study due to their long-lived electronic spin lifetimes \cite{kikkawa1998}.

To our knowledge, no theoretical study has been carried out so far on the effect of the use of different Born approximations for the impurity screening on electronic spin relaxation. In this work we aim to perform this analysis for bulk {\it n}-type GaAs subject to DP mechanism. We calculated the spin relaxation times within LTFA in B1 and B2 by ensemble Monte Carlo simulations. Our findings show that spin relaxation times are reduced by a substantial amount, of the order of hundred(s) of picoseconds,  when calculated within LTFA in B2 approximation. In addition, for lattice temperatures $T > 50$ $\mathrm{K}$,  $T<T_F$ ($T_F$ the Fermi temperature),   and low to intermediate carrier densities, we find that our numerical SRT results show the  temperature dependence behaviour predicted by the Dyakonov-Perel theory when including electron-impurity scattering only.  This supports that electron-impurity interactions are dominant at those temperatures in bulk GaAs.

This paper is organized as follows. In Sections \ref{sec:1}  and  \ref{sec:2}   we recall the main results of finite temperature Thomas-Fermi
screening theory in different Born approximations and the computational model employed to study the spin dynamics in GaAs.
In  Section \ref{sec:4} we present and discuss our results for spin relaxation times in first and second Born approximation for a range of temperatures and given doping densities for which we expect our physical approximations, i.e. RPA and Boltzmann statistics, to be reliable. Finally Sec. V summarizes our conclusions.

\section{Inverse Screening Length in different Born Approximations}\label{sec:1}

Within RPA and LTFA the Coulomb interaction potential between an impurity of charge $Z$ and an electron in the conduction band (CB) is given by

\begin{equation}\label{eq:potential}
  V\left(r\right)=-\frac{Z e^{2}}{4 \pi \varepsilon r}
  \mathrm{e}^{-\beta_{\mathrm{TF}} r } \, ,
\end{equation}
where $r=|\mathbf{r}_{A}-\mathbf{r}_{B}|$ is the distance between the   carrier $A$ and impurity $B$ at coordinates
$\mathbf{r}_{A}$ and $\mathbf{r}_{B}$ respectively,  $\varepsilon$ is the material dielectric constant,
$\varepsilon=12.9 \, \varepsilon_{0}$ for GaAs, and $\varepsilon_{0}$ is  the vacuum permittivity. The quantity
$\beta_{\mathrm{TF}}$ is called Thomas-Fermi wavevector or \textquotedblleft inverse screening length\textquotedblright .
Note that throughout this work we shall consider only single charge impurities, i.e. $Z =1$.

In  semiconductors  the inverse screening length can be obtained exploiting the Friedel sum rule (FSR). The FSR which holds for Fermi liquids \cite{langer1961} and free Fermi gas as well \cite{mahan2000}, states that the impurity charge must be completely screened at finite distance by  itinerant carriers \cite{friedel1958}.
The Friedel Sum Rule for a {\it n}-type semiconductor with one parabolic band, according to Stern's formula \cite{stern1967} is
\begin{equation}\label{eq:fsr}
 \frac{2}{\pi}\sum_{l=0} ^{\infty}\left(l+1 \right)\int_{0}^{\infty} f_{FD}\left(E\right) \frac{d\delta_l\left(E\right)}{d E} d\, E= Z \, ,
\end{equation}
where  $f_{FD}$ is Fermi-Dirac distribution and    $\delta_l$ are the phase shifts  \footnote{In the
degenerate case ($T=0$ $\mathrm{K}$) the phase shifts must be evaluated at Fermi energy $E_\mathrm{F}$ or equivalently at Fermi wavector $k_\mathrm{F}$.} due to the presence
of the electron-impurity potential (given by Eq. \ref{eq:potential} in our case) relative to the $l$ -th partial wave solution of
the Schr\"{o}dinger radial equation for  angular momentum numbers $l=0,1, \, \cdots$. Indeed the phase shifts $\delta_l$ are then constrained by FSR.

In B1 approximation the phase shifts $\delta_l$ are given by \cite{joachain1987}
\begin{equation} \label{eq:B1shifts}
\left(\tan \delta_l \right)_{\mathrm{B1}} = - \frac{2 m^{\ast} k}{\hbar^{2}} \int_{0}^{\infty} j_l^{2}\left(kr\right) V\left(r\right) r^{2} d\, r \, ,
\end{equation}
where $k$ is the wavevector magnitude of the colliding carrier of effective mass $m^{\ast}$ and energy $E= \hbar^{2} k^{2}/2 m^{\ast}$ and $j_l$ are the spherical Bessel functions.  The bottom of the $\mathrm{\Gamma}$ valley of GaAs as single ideal parabolic energy  band corresponds to a carrier effective mass $m^{\ast}=0.067 \, m_{\mathrm{e}}$ where $m_{\mathrm{e}}$ is the electron bare  mass \cite{vurgaftman2001}.

 If $B1$ approximation holds it is usually assumed that the $\delta_l$ are small, and then $\left(\tan \delta_l \right)_{\mathrm{B1}} \approx\delta_l \equiv
\delta_{l,\mathrm{B1}}$. Furthermore when B1 holds, then the phase shifts in B2 are smaller than the ones
calculated in B1, i.e., $\delta_{l,B2} < \delta_{l,B1}$ \cite{nersisyan2013}\footnote{It is worthwhile to recall here that in general if the true and the first Born approximation phase shifts are small this indeed does not imply the validity of the Born approximation for a general short-range potential \cite{peierls1979}.}.

Assuming that B1 holds, inserting Eqs. \ref{eq:potential}, \ref{eq:B1shifts} in Eq. \ref{eq:fsr} and using
the identity $\sum_{l} \left(2l+1\right) j_l^{2} =1 $, one obtains the following expression for the
inverse screening length $ \beta_{\mathrm{TF}}$, i.e. $\beta_{\mathrm{TF,B1}} \equiv \beta_{\mathrm{B1}}$ \cite{chattopadhyay1981,stern1967}

\begin{equation}\label{eq:screening}
\beta_{\mathrm{B1}}^{2}=\frac{n_{\mathrm{e}} e^{2}}{\varepsilon k_{\mathrm{B}}T}\frac{\mathscr{F}_{-1/2}(\eta)}{\mathscr{F}_{1/2}(\eta)} \, .
\end{equation}
Here  $n_{\mathrm{e}}$ is the electronic density, $k_{\mathrm{B}}$  is the Boltzmann constant, $T$ is the lattice temperature,
 $\mathscr{F}_{j}$ denotes the Fermi-Dirac integral of order $j$  \cite{blakemore1962} and finally $\eta=\mu/\left(k_{\mathrm{B}}\mathrm{T}\right)$
is the reduced electronic chemical potential. Eq. \ref{eq:screening} is often referred to as 'Dingle's theory of screening'   \cite{sanborn1995}.
Dingle's theory is indeed equivalent to a finite temperature linearized Thomas-Fermi approximation and manifestly disregards the electron-electron
exchange and correlation effects \cite{dingle1955}. This is clearly consistent with RPA.

In order to find out the inverse screening length in B2 one can use the Schwinger variational principle for the phase shifts.
With the trial function $u_l\left(r\right) = r j_l \left(kr\right)$  this principle provides the useful relations \cite{joachain1987}

\begin{equation}\label{eq:schwinger}
\tan \delta_l =- k A_l \left(1-B_{l}/A_{l} \right) \, ,
\end{equation}
and
\begin{equation}\label{eq:Al}
A_l=  \int_{0}^{\infty} j_l^{2} \left(kr \right) U\left(r \right) r^{2} dr \, ,
\end{equation}
and
\begin{align}\label{eq:Bl}
 B_l &= \int_{0}^{\infty} d r  \int_{0}^{\infty} d r' j_l\left(kr\right) V\left(r\right) G_l \left(r,r'\right) \nonumber \\
 & \times U\left(r'\right) j_l\left(kr'\right) r^{2}r'^{2} \, .
\end{align}
In Eqs. \ref{eq:Al} and \ref{eq:Bl}, we have introduced the usual reduced potential $U\left(r\right)= 2 m^{\ast}V\left(r\right)/\hbar^{2}$ and the following function $G_l$
\begin{equation}
 G_l\left(r, r' \right) = k j_l \left(k r_{<} \right) \eta_l  \left(k r_{>} \right) \, ,
\end{equation}
where $\eta_l$ are the spherical Neumann functions,  $r_{<} = \min \{r,r'\}$, and $r_{>} = \max \{r,r'\}$.

Eq. \ref{eq:schwinger} corresponds to the Born series through second order \cite{joachain1987}. Again we shall assume
that we can approximate $\tan \delta_l$ by $\delta_l$ in Eq.  \ref{eq:schwinger} to the second order. Moreover notice
that,  to first order ($B_l=0$), Eq. \ref{eq:schwinger} gives the B1 phase shifts $\delta_l = -k A_l$, the same
of Eq. \ref{eq:B1shifts}.
In the limit of low temperatures (see Section \ref{sec:4} for the related discussion),  and assuming that only $s$-waves ($l=0$) matters for corrections to the first order,  one then can write Schwinger variational principle as
\cite{patterson1989}
\begin{equation}\label{eq:schwinger1}
 \delta_l \simeq -k A_l\left(1-B_0/A_0\right) \,.
\end{equation}

Finally Patterson and Lehoczky, by assuming that only for $kr \ll 1$ there are important contributions to the term $B_0/A_0$ in Eq. \ref{eq:schwinger1},
obtained the following formula for  the inverse screening length $\beta_{\mathrm{B2}} $ in B2   \cite{patterson1989}

\begin{equation}
 \beta_{\mathrm{B2}} = C\left(\beta_{\mathrm{B1}}\right) \beta_{\mathrm{B1}} \, ,
\end{equation}
where we defined the function $C$ of variable $\beta_{\mathrm{B1}}$  by
\begin{equation}
 C\left(\beta_{\mathrm{B1}}\right) = \frac{\beta_{\mathrm{B1}}}{M+\sqrt[]{M^{2}+ \beta_{\mathrm{B1}}^{2}}} \, ,
\end{equation}
and the negative constant $M$ as
\begin{equation}
 M = -\frac{m^{\ast} Z e^{2}}{8 \pi \epsilon \hbar^{2}} \, .
\end{equation}

\begin{figure}
\resizebox{0.55\textwidth}{!}{%
  \includegraphics{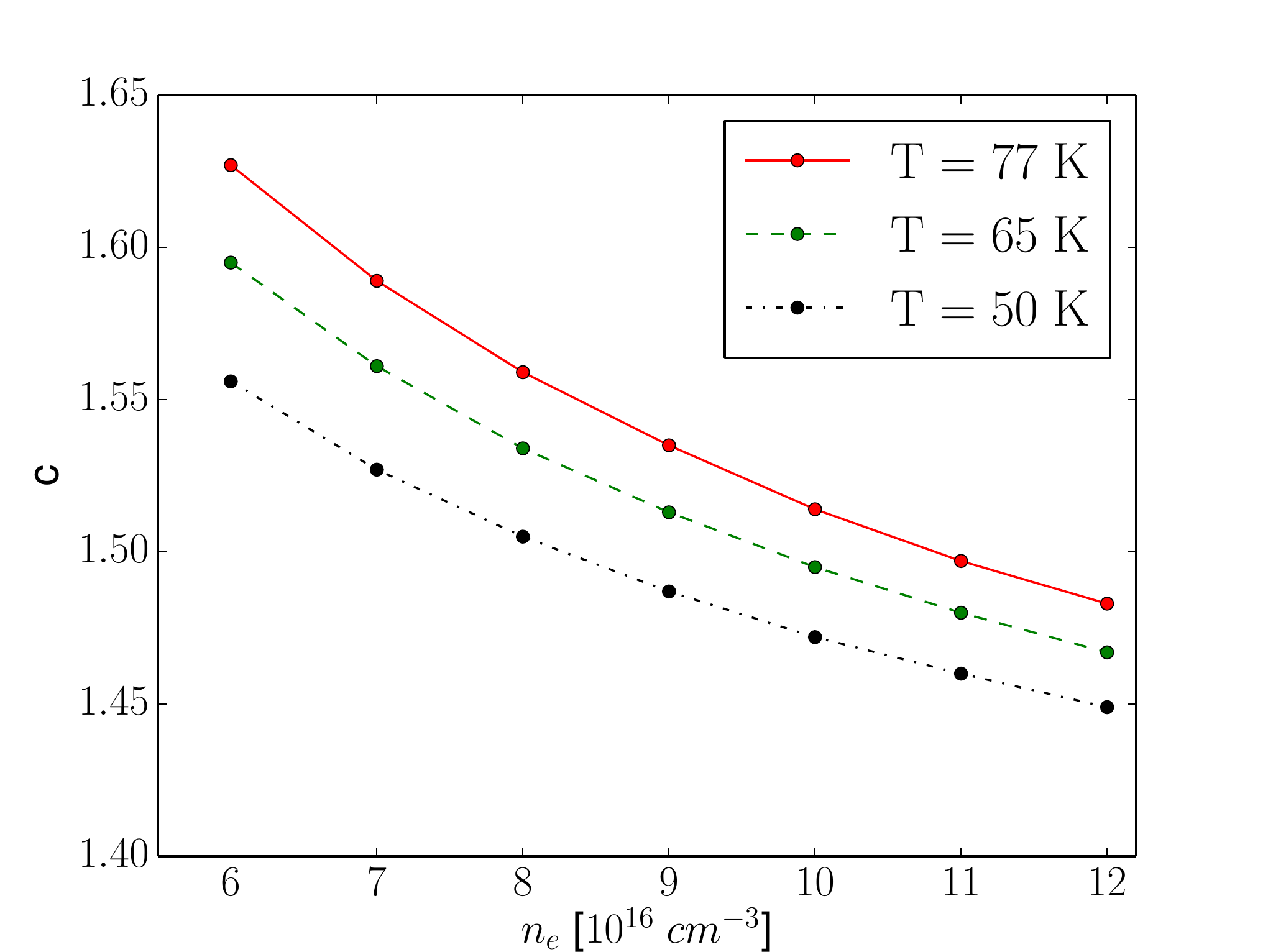}
}
\caption{Variation of the function C upon the temperatures
and electron densities of interest.}
\label{fig:factorC}       
\end{figure}

In Fig. \ref{fig:factorC} we plot the function $C$ dependence for a range of temperatures and electronic densities.
From Fig. \ref{fig:factorC} it is evident that, by increasing the electronic densities, the dependence of  $C$  on temperature becomes less important, as  we go from 5\% spread with temperature for the smaller density to 3\% for the largest density considered. Indeed at high electronic densities and for the temperatures considered in \ref{fig:factorC}, $C$ becomes temperature independent, e.g. $C = 1.27$ for $n_e = 10^{18} ~cm ^{-3}$ regardless the temperatures of interest.
This is not a surprise as the increase of the impurity concentration $n_i$ gives a more metallic character to the semiconductor. Note that we shall assume full ionization throughout,  i.e.  $n_{\mathrm{i}}=n_{\mathrm{e}}$  according to Ref. \cite{jiang2009}.

More importantly we observe that in the case of donor impurity always $\beta_{\mathrm{B2}}> \beta_{\mathrm{B1}}$ which means that
the range of the interaction potential, roughly $\beta_{\mathrm{TF}}^{-1}$, become shorter when screening is accounted in B2, and hence the relative electron-impurity (e-i) scattering probability becomes smaller \cite{joachain1987}.

\section{Physical Model and Computational Method}\label{sec:2}

In this paper we consider Ensemble Monte Carlo simulations of electronic spin transport in 3-dimensional GaAs. In the following
we briefly recall the details of the physical model and of the computational method;
for more details, we refer the reader to our previous works \cite{marchetti2014,marchetti2014a} and the references therein.

Within a parabolic approximation, representing the bottom of the central $\mathrm{\Gamma}$ valley,
the carriers undergo collisions  with lattice excitations (longitudinal acoustic (LA)  phonons,  polar
longitudinal optical (LO) phonons) and singly-ionized impurities according to Brooks-Herring (B-H) model \cite{jacoboni1989}.
Phonons are considered at equilibrium at the lattice temperature $T$.
Electron-electron scattering is accurately implemented for a nondegenerate regime, which means that
the carriers are simulated  as distinguishable classical particles and
their scattering time $\tau_{\mathrm{ee}}$ is then proportional to the electron density, $\tau_{ee} \sim n_{\mathrm{e}}^{-1}$.
In the present work we simulated $N=25,000$ carriers.
We note in passing that for
the case of a three-dimensional interacting electron gas in the parameter range considered the number of exchange processes is negligible \cite{giuliani2005}.
All the scattering rates are calculated through  Fermi's Golden rule and implemented within Ensemble Monte Carlo (EMC) method \cite{jacoboni1989}.
The Ensemble Monte Carlo method solves numerically the Boltzmann equation for charge transport and therefore determines the carriers
free flight times and their scattering events for each simulated particle \cite{jacoboni1989}.  EMC simulations are performed until
enough data are generated according to the aims of the study. The carrier's  spin degree of freedom can be included in EMC simulations
either using  density matrix formalism or full spinor wavefunction. We choose the latter approach in which each spinor wavefunction $\psi$ is acted
upon by an unitary time-evolution operator $\hat{U}\left(t\right)=\exp(-iH_{\mathrm{D}}t/\hbar)$ generated by the Dresselhaus Hamiltonian  $H_{\mathrm{D}}$
\cite{fabian2007}
\begin{equation}\label{eq:dresselhaus}
H_{\mathrm{D}}=\hbar  \mathbf{\Omega}(\mathbf{k})\cdot \vec{\mathbf{\sigma}} \, ,
\end{equation}
where $\vec{\mathbf{\sigma}}=\left(\sigma_x,\sigma_y,\sigma_z\right)$ are the Pauli matrices, and the Larmor precession frequency vector $\mathbf{\Omega}(\mathbf{k})$ is
\begin{equation}\label{eq:larmor}
 \mathbf{\Omega}(\mathbf{k})=\frac{\gamma_{so}}{\hbar}[k_{x}(k_{y}^2-k_{z}^2), k_{y}(k_{z}^2-k_{x}^2), k_{z}(k_{x}^2-k_{y}^2)] \, .
\end{equation}
Here $k_{i}$  are the wavevector components along the cubic crystal axes, $i=x,y,z$, and  $\gamma_{so}$ is the spin-orbit coupling (SOC), also called
Dresselhaus coefficient. In the present work we shall assume $\gamma_{so}= 21.9$ $\mathrm{eV}$ \AA $^{3}$ according to \cite{marchetti2014}.
The single particle spinor wavefunction $\psi$ at some later time $t > t_0=0$ is then given by
\begin{equation}\label{eq:spinEvol}
 \psi\left(t\right) = \hat{U}\left( t\right) = \mathrm{exp}(-iH_{\mathrm{D}} t/\hbar)\psi\left(t_0\right) \, .
\end{equation}

After the system relaxes to thermal equilibrium, typically in a few picoseconds, we set the electron spins along one direction, namely the $z$-axis.
Their time-evolution is then dictated by Eq. \ref{eq:spinEvol} causing  spin dephasing.  At any given time we can extract the
expectation values of the $S_x$, $S_y$ and $S_z$ components of the individual electron spin operator $\hat{S}$ to get the
probability for the spin to be aligned along each direction. Because we  start from an electronic ensemble fully polarized along the $z$ -axis,
we focus on the time evolution of the expectation value of the total  $z$-component spin operator $\hat{S}_{z,\mathrm{tot}}$. For each simulation, by plotting
$S_{z,\mathrm{tot}}$ against time, and assuming an exponential decay, we fit  the  data from the simulation
and extract the corresponding spin relaxation time $\tau_\mathrm{s}$ \cite{marchetti2014}.

\subsection{The $(n_{\mathrm{e}},T)$ Plane} \label{sec:3}

\begin{figure}
\resizebox{0.50\textwidth}{!}{%
  \includegraphics{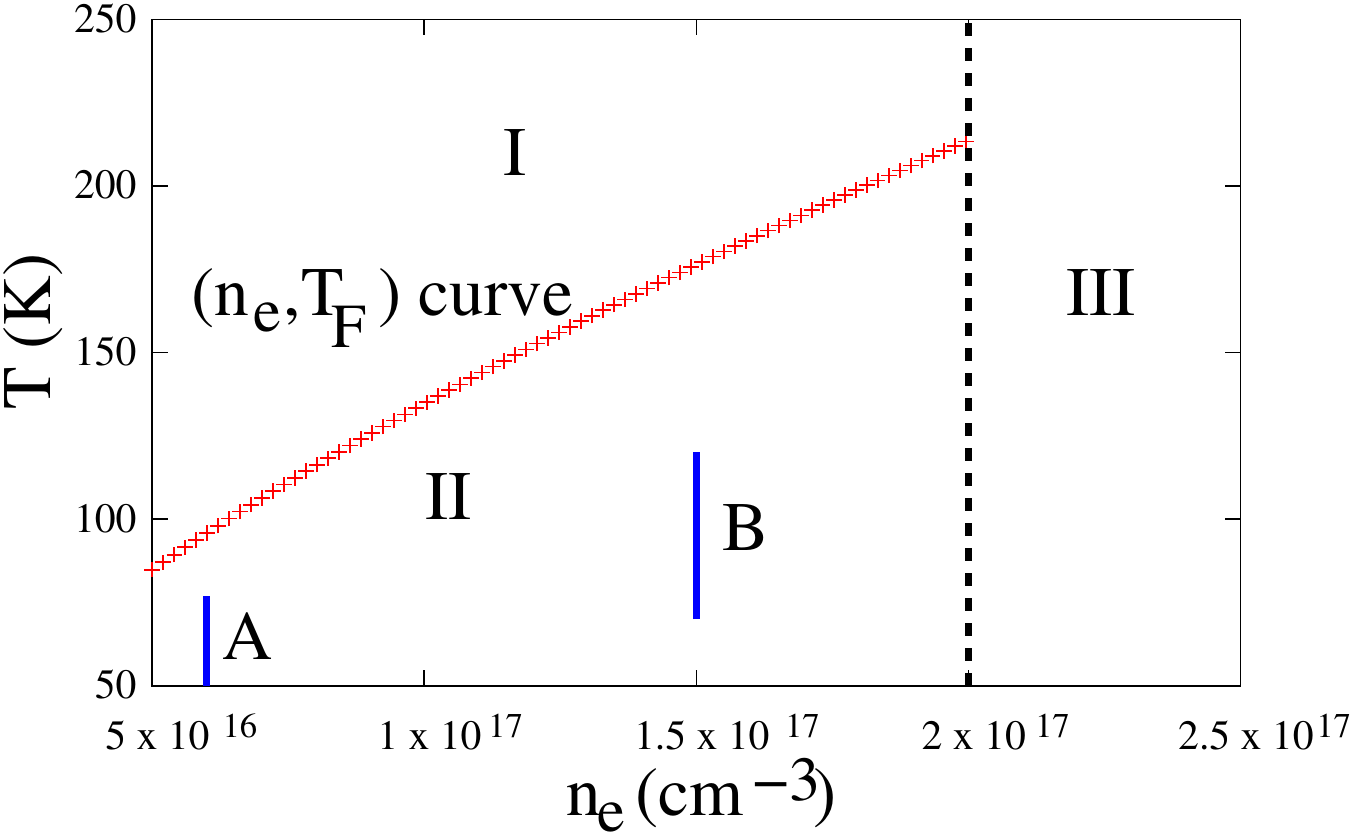}
}
\caption{Parameter plane $(n_e, T)$ for GaAs. Spin relaxation times are
calculated along the blue lines $A$ and $B$. The red curve indicates the $T_F (n_e)$ curve. The vertical dashed line corresponds to $r_s=1$. In regions I and II $1.0\le r_s\le 1.7$. }
\label{fig:region}       
\end{figure}

In order to perform  calculations consistent with the physics discussed so far,
we need to find a region on the plane $(n_{\mathrm{e}},T)$  where at the same time RPA holds
and Dyakonov-Perel is the dominant spin relaxation mechanism.
Moreover this region should include low/intermediate temperatures in a way that Patterson and Lehoczky approximation (PLA) is expected to be valid.  Fig. \ref{fig:region} shows a partition of plane $(n_{\mathrm{e}},T)$ where the red curve
corresponds to the points $(n_e, T_{\mathrm{F}})$, being $T_{\mathrm{F}}$ the Fermi temperature relative to the electron
density $n_e$. Its end points are $(n_1 = 5 \times 10^{16}, T_{\mathrm{F_1}}= 85)$, corresponding to a Wigner- Seitz radius $r_s=1.7$, and $(n_2 = 2\times  10^{17}, T_{\mathrm{F_2}}= 215)$, corresponding to $r_s=1.0$,
where densities and temperatures are assumed in $\mathrm{cm}^{-3}$ and $\mathrm{K}$ units respectively.
We note that for densities smaller than $n_1 = 5 \times 10^{16}$  $\mathrm{cm}^{16}$  and temperatures smaller than 50 $\mathrm{K}$, GaAs behaves
like an insulator (electrons are localized at donor sites), while in region $III$ electron-plasmon interaction may become important \cite{jacoboni1989}.
In regions $I$ and $II$, see Fig.~\ref{fig:region}, RPA holds because the  Wigner- Seitz radius $r_s \simeq 1$ \cite{giuliani2005}, and Elliot-Yafet spin relaxation
is less important than Dyakonov-Perel mechanism \cite{jiang2009}. For temperatures $T\ll T_{F}$ the pure  quantum behaviour of the electron gas is important. In this case the tiny electron-electron cross-section stems directly from Pauli principle. However when working in intermediate regimes, $T\stackrel{<}{\sim} T_{F}$, due to the negligible number of exchange processes in a three-dimensional interacting electron
gas \cite{giuliani2005}, an electron-electron scattering which includes only direct processes can be suitable for carriers'
dynamics simulation, at the same time ensuring that the system thermalizes properly.
We will then calculate spin relaxation times along the solid lines $A$ and $B$, see Sect.~\ref{sec:4}.

\section{Results and Discussion}\label{sec:4}

Patterson and Lehoczky assumed that their correction to the inverse screening in B2  holds at low temperatures without giving any specific information about the temperature range.
Clearly this correction makes sense insofar  B1 approximation for electron-impurity scattering fails, which from previous studies it is known to happen for temperatures  $T \approx 5 \div 80$ $\mathrm{K}$ in semiconductors like GaAs and Si, see Refs.~\onlinecite{meyer1982},\onlinecite{meyer1981}.

\begin{figure}
\resizebox{0.55\textwidth}{!}{%
  \includegraphics{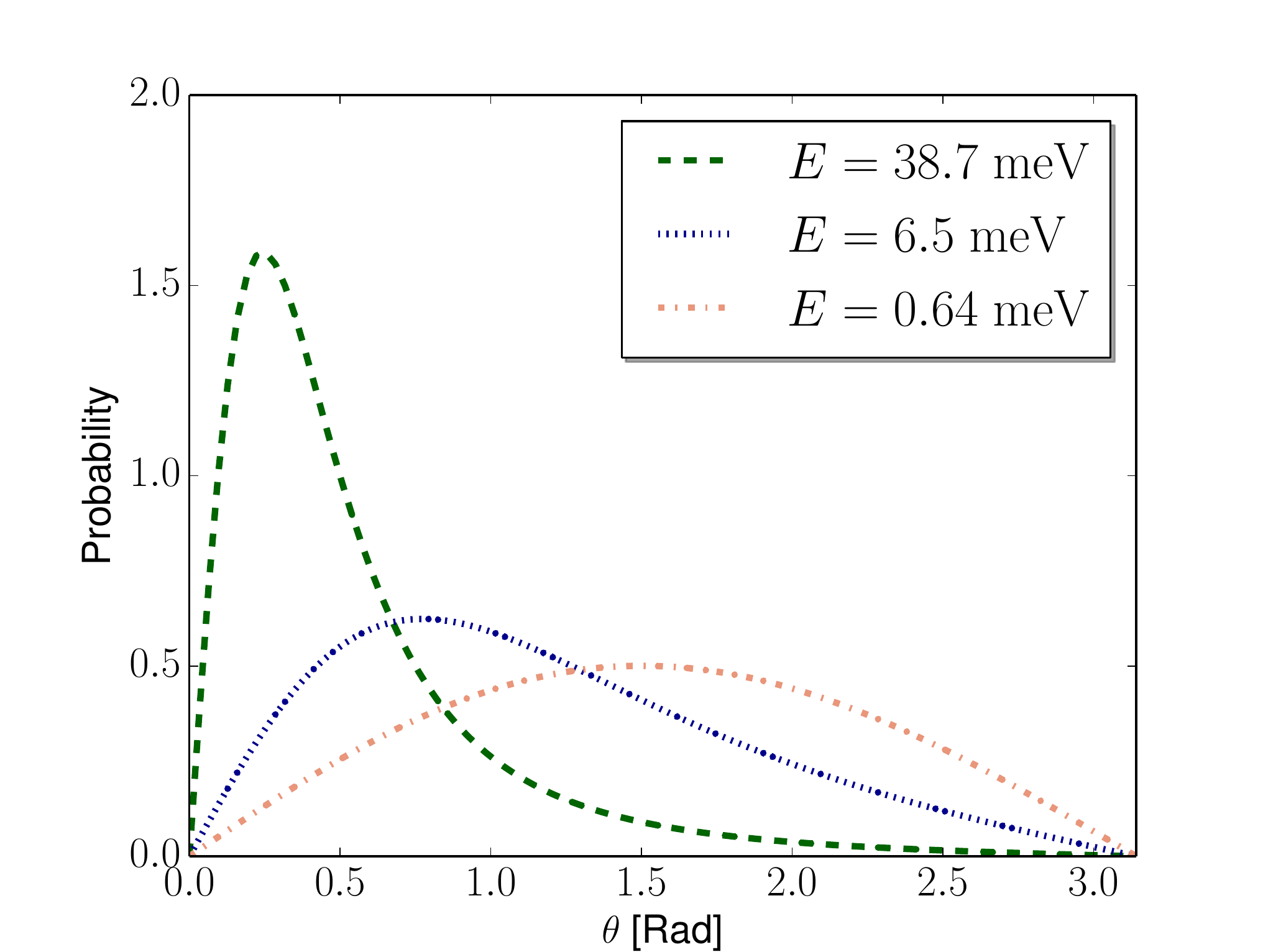}
}
\caption{ Normalized angular probability in B-H approach against e-i scattering angles in radiants for different carrier's energy $E=38.7, 6.5, 0.64$
$\mathrm{meV}$ corresponding to temperatures $T=300, 50, 5$ $\mathrm{K}$ respectively. The inverse screening
length $\beta_{\mathrm{B1}}$ is calculated assuming $n_e = 6 \times 10^{16}$ $cm^{-3}$.}
\label{fig:probability}       
\end{figure}

In this work we are simulating the system dynamics using EMC, which means that we are in the privileged position of been able to observe and characterize the scattering
events directly, e.g. tracking the overall occurrence of a certain type of scattering or its momentum distribution,  and we can then infer from this analysis the validity range of B1.
More precisely due to the large number of collisions in  EMC simulations, we may expect that PLA corrections to B2  to be important whenever the majority of e-i collisional events happens  in s-wave  or in p-wave ($l=1$) at very low energy \footnote{In a semiclassical picture the substantial contribution to e-i scattering comes from waves for which $l\leq kb$ where $b$ is the impact parameter and $k$ the electron wavevector. \cite{joachain1987}. For instance, assuming that $b$ is half the interatomic distance between the impurities and $k\simeq k_F$, the Fermi momentum, one can see that only few waves in general matter for slow electrons.}.
 To this end, in  Fig.~\ref{fig:probability} we plotted the (normalized) angular probability for e-i scattering according to B-H model \footnote{ To date, Brooks-Herring approach is one of most reliable and successful physical picture of electron-ion interaction. The other widely used model, by Conwell and Weisskopf  \cite{chattopadhyay1981}, is instead based upon the unscreened Coulomb which is inconsistent with RPA.} employed in our EMC code, for different electron thermal energies $E =3/2 k_{\mathrm{B}} T$ (classical gas) corresponding to different lattice temperatures \footnote{ Note that from the kinematics of non-relativistic collisions one can show that the scattering angle $\theta$  in laboratory  frame of reference is very close to the corresponding scattering angle  $\theta_{CoM}$ in centre of mass frame of reference, i.e. $\theta \approx \theta_{CoM}$ because $m^{\ast}\ll M_I$, being $M_I$ the mass of impurity centers \cite{kamal2014}  Moreover the maximum scattering angle $\theta_{max}$ is obtained when $\theta =\pi$ \cite{kamal2014}.}.

On this premise we have calculated the spin relaxation times $\tau_s$ due to DP mechanism along two curves in region $II$, lines $A$ and $B$, see Fig.~\ref{fig:region}.
Line $A$ corresponds to  $n_e = 6 \times 10^{16}$ $cm^{-3}$  ($T_{\mathrm{F}} = 85.7$ $\mathrm{K}$) and  $T = 50 \div 77$ $\mathrm{K}$. The reasons for this choice are the following: a low doping density is consistent with an impurity single-site model
due to large interatomic distances and then should not require a multi-ion screening correction to the linearized Thomas-Fermi screening theory of Section \ref{sec:1}.  Moreover an intermediate regime is provided by the chosen temperatures which are smaller, but not  much smaller, than $T_{\mathrm{F}}$.  Furthermore we refrain from going to lower temperatures, i.e. $T < 50$  $\mathrm{K}$,  because the collisions of carriers with LA phonons become inelastic, a case which is not included in our calculations. The only inelastic processes in our EMC simulations are due to LO phonons absorption and emission.

\begin{figure}
\resizebox{0.50\textwidth}{!}{%
  \includegraphics{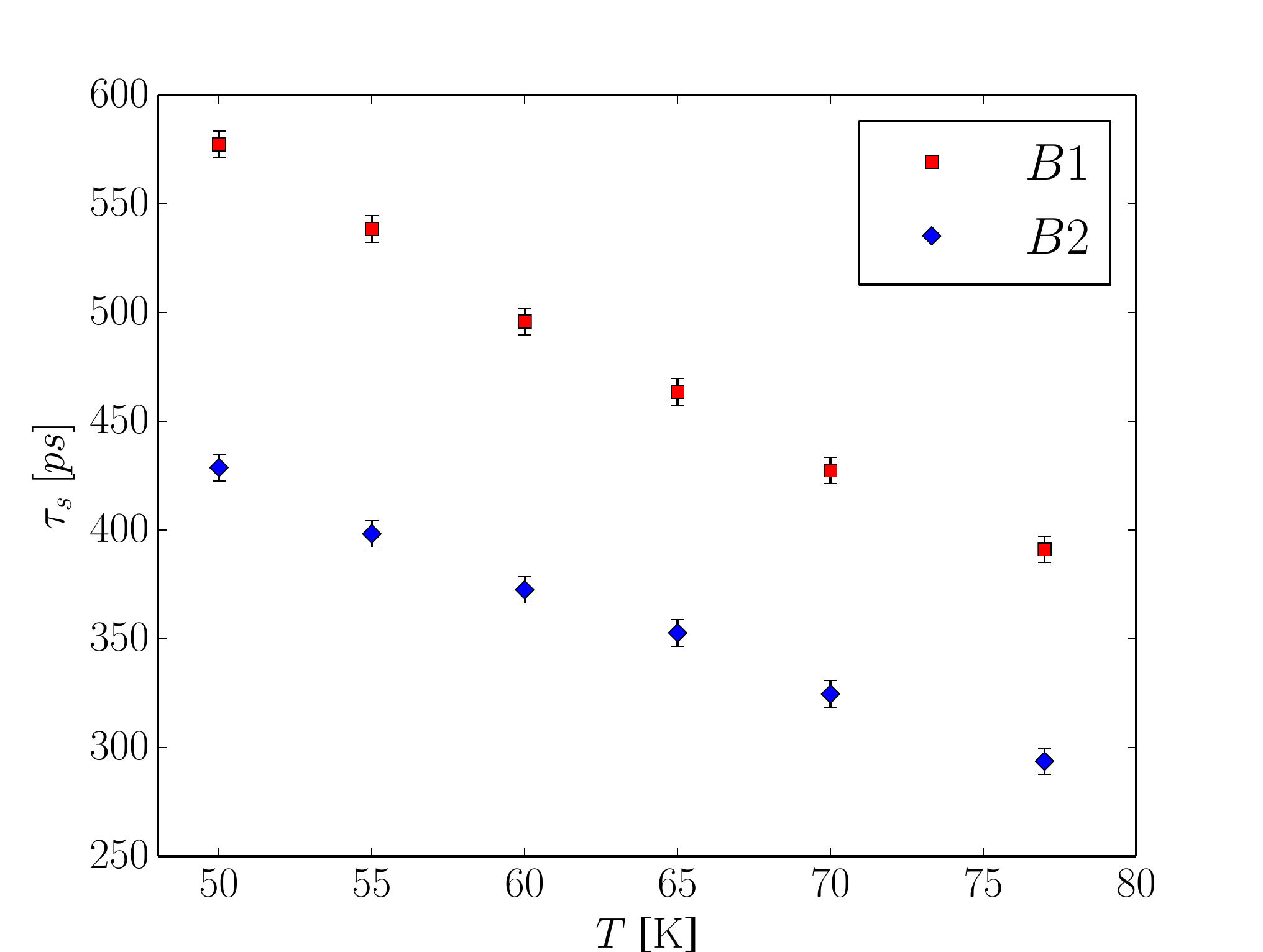}
}
\caption{Comparison of the spin relaxation times $\tau_s$ along line $A$ for $B1$ (squares) and $B2$ (diamonds)
approximations. Here $n_{\mathrm{e}}=
6\times 10^{16}  \mathrm{cm}^{-3}$, $N= 25,000$ and $\gamma_{so}= 21.9 \, \mathrm{eV} \angstrom ^{3}$.}
\label{fig:SRTcurve}       
\end{figure}

The computed SRT curves along line $A$ for the two Born approximations of the  screening length are shown in Fig.~\ref{fig:SRTcurve}. They both show a monotonic temperature dependence, with the SRT increasing for decreasing temperatures. However, while qualitatively the behavior of the curves is similar, quantitatively the correction due to B2 is very substantial: B1 predictions overshoot the B2 curve by 35\% at $T=50$ $\mathrm{K}$ and by 34\% at $T=77$ $\mathrm{K}$.

 The increase of SRT with temperature can be  explained  in terms of spin precession about randomly fluctuating magnetic fields and temperature dependence of electron-impurity scattering.
Decreasing the temperature, the B-H electron-impurity  scattering rate $\tau_{ei}^{-1}$ increases. This is consistent with the e-i
contribution to electron mobility in GaAs. Indeed Hall mobility curves show that for $T = 5 \div 100$ $\mathrm{K}$
 the electron-impurity scattering mainly controls the mobility \cite{chattopadhyay1981}.

The total scattering rate in the system is given by $\Gamma_{tot} = \tau_{ei}^{-1} + \tau_{ee}^{-1}+ \tau_{ap}^{-1}+\tau_{op}^{-1}$ where
the last two terms denote LA and LO phonon scattering rates respectively \footnote{The EMC algorithm requires that the carriers also undergo  to a self-scattering where nothing actually happens \cite{jacoboni1989}.}.
Then the Dyakonov-Perel spin decoherence mechanism is characterized by \cite{zutic2004}
\begin{equation}
 \frac{1}{\tau_s} = \frac{\Omega_{av}^{2}}{ \Gamma_{tot}} \, ,
\end{equation}
where $\Omega_{av}$ is the average magnitude of $\mathbf{\Omega}(\mathbf{k})$, see Eq.~\ref{eq:larmor}, over the momentum distribution.
When lattice temperatures decrease, $\Gamma_{tot}$ increases due to e-i collisional rates mainly.  Indeed, from our EMC simulations performed at $T= 60 $ and then at $ T= 50$  $\mathrm{K}$ for $n_{\mathrm{e}}=
6\times 10^{16}  \mathrm{cm}^{-3}$ we note an overall increase of the number of
Coulomb (e-e plus e-i) collisions  of about $5$ \% which indeed slows the spin dephasing.

According to the previous physical picture we expect that $\tau_s^{B2} < \tau_s^{B1}$, where $\tau_s^{B1}, \tau_s^{B2}$ to be spin relaxation times in B1 and B2  respectively: in fact the screening computed in $B2$ reduces the e-i scattering probability compared to the other scattering mechanisms, and this reduction is correctly simulated by the EMC algorithm.  In Fig.~ \ref{fig:SRTcurve} the curves
for $\tau_s^{B1}$ (squares) and $\tau_s^{B2}$ (diamonds) show indeed the correct behaviour. As noted, $\Delta \tau_s = \tau_s^{B1}-\tau_s^{B2}$,
is very substantial, and becomes larger as  the temperature decreases, where also one expects that the Patterson and Lehoczky approximation becomes more accurate.

\begin{figure}
\resizebox{0.55\textwidth}{!}{%
  \includegraphics{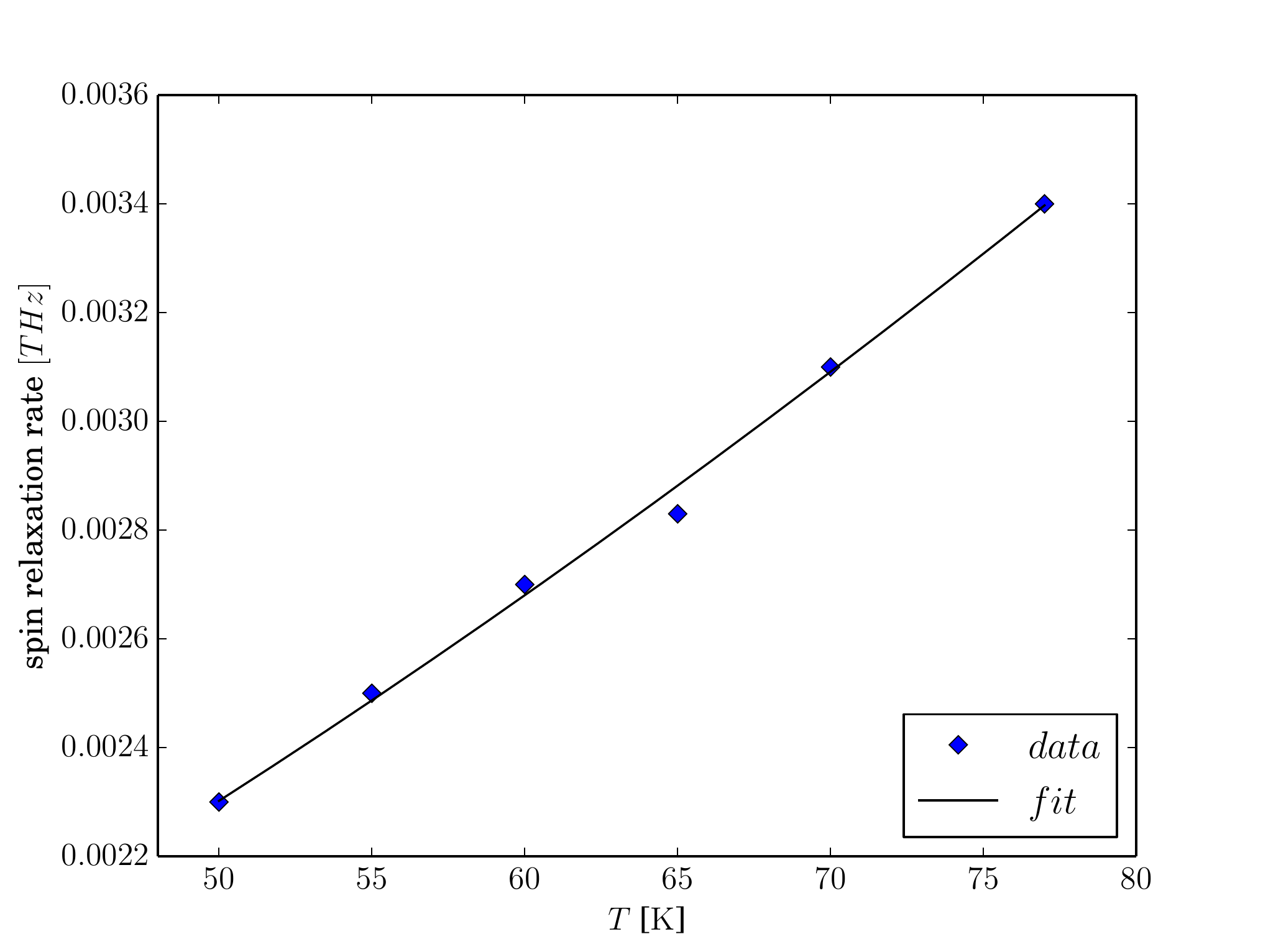}
}
\caption{Fit of spin relaxation rate data for curve A (diamonds) by the curve $\Gamma_s =a+bT^{3/2}$  ($a=3.4 \times 10^{-6}$, $b=0.0011$).}
\label{fig:fit}       
\end{figure}

The temperature dependence of the DP spin relaxation rate due to charge impurity scattering $\Gamma_{s,e-i}$  is expected to be $\Gamma_{s,e-i} \sim T^{3/2}$ \cite{zutic2004}. The fitting of our results for the spin relaxation rate, $\tau_s^{B2}$,  shows a very similar temperature dependence, see Fig.~\ref{fig:fit}. This supports that e-i scattering is the dominant scattering along line $A$. We obtain similar results for curve $B$ (not shown).

We can now further investigate whether or not the second Born approximation is adequate to our problem. A sufficient condition for the convergence of Born series at all energies is given by \cite{joachain1987}
\begin{equation}\label{eq:suff_cond}
I = \frac{2 m^{\ast} }{\hbar^{2}} \int_{0}^{\infty} d r \, r \left|V\left(r\right)\right| < 1 \, ,
\end{equation}
which, together with one of the several necessary conditions for the existence of at least one
bound state \cite{calogero1965}, says that the convergence happens when the potential $V$ does not support any bound state \cite{joachain1987}. Kohn has shown that for very low energy collisions the Born series converges
if \cite{kohn1954,joachain1987}:
\begin{equation}\label{eq:suff_cond1}
 I < 2l + 1 \, .
\end{equation}

\begin{figure}
\resizebox{0.55\textwidth}{!}{%
  \includegraphics{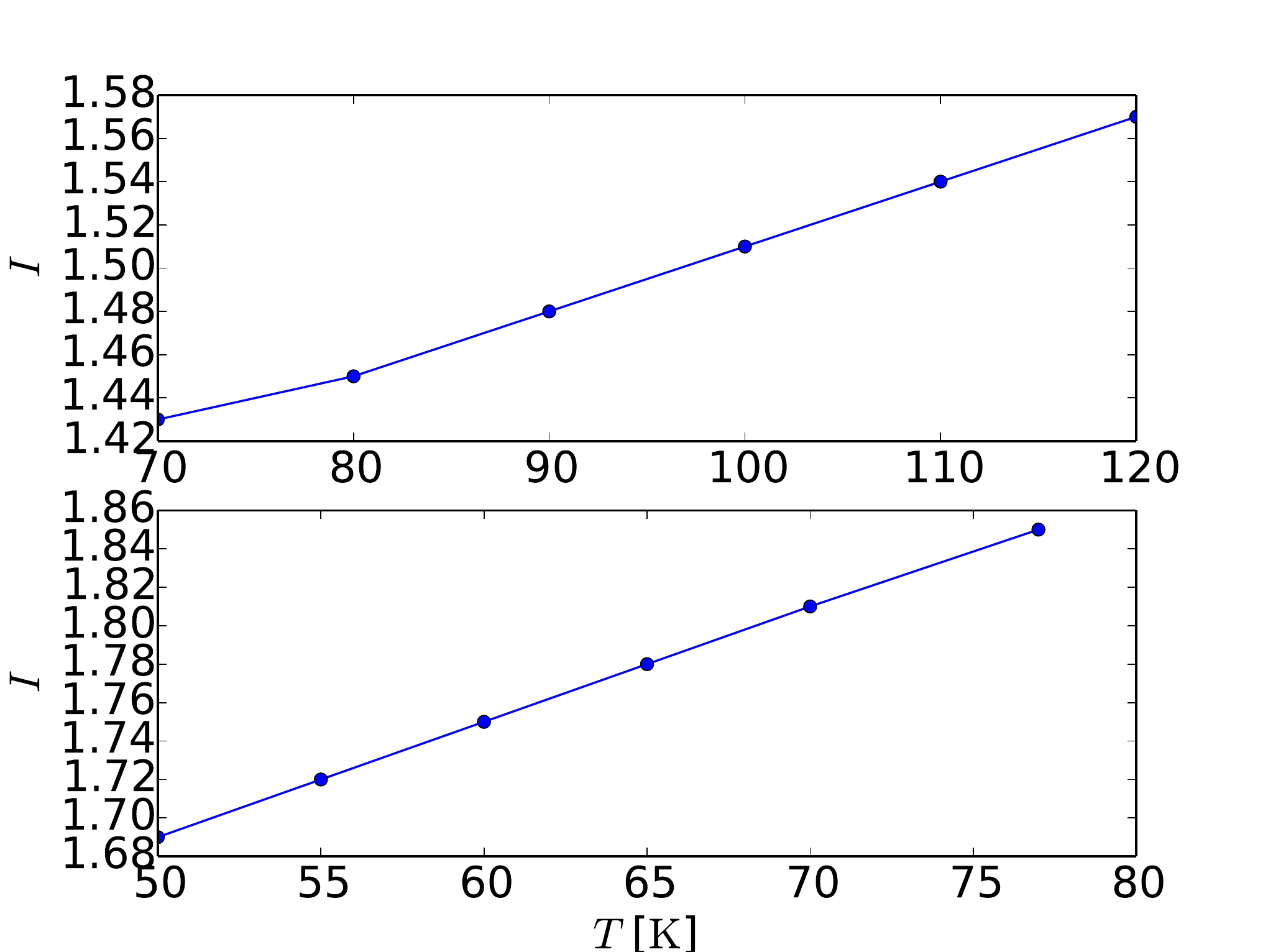}
}
\caption{Values of integral $I$ against temperatures along line $A$ (bottom panel) and line $B$ (top panel).}
\label{fig:upperBound}       
\end{figure}

Inserting Eq.~\ref{eq:potential} with $\beta_{\mathrm{B1}}$ in the expression for $I$ from Eq.~\ref{eq:suff_cond}, we obtain
\begin{equation}
 I = \frac{2 m^{\ast} }{\hbar^{2}} \frac{Ze^{2}}{4\pi \epsilon} \frac{1}{\beta_{\mathrm{B1}}} \, ,
\end{equation}
which becomes
\begin{equation}
 I = \frac{2 Z}{a^\ast_{\mathrm{B}}\beta_{\mathrm{B1}}} \, ,
\end{equation}
where we used the effective Bohr radius $a^\ast_{\mathrm{B}}=(4\pi\hbar^{2} \varepsilon )/(\mathrm{e}^{2}m^{\ast})$. We plot
$I$ in Fig.~ \ref{fig:upperBound} for $n_e = 6 \times 10^{16}$ $cm^{-3}$ and  $T =30 \div 77$ $\mathrm{K}$ (related to curve $A$) and $n_e = 1.5 \times 10^{17}$ $cm^{-3}$ and  $T =70 \div 120$ $\mathrm{K}$ (related to curve $B$). Fig.~ \ref{fig:upperBound} indeed shows that  the inequality Eq.~\ref{eq:suff_cond1} is satisfied for $\beta_{\mathrm{B1}}$ (and hence for $\beta_{\mathrm{B2}}$) for $l=1$, and thus the Born series should converge for the range of temperatures and densities explored. In this regard, the results in B2 should then be more accurate than the corresponding ones, computed in B1 insofar the PLA holds.
We note that  the condition for validity of B1 along $A$ and $B$ curves is not satisfied at low energies, as it would required that $I/2 \ll 1$ \cite{marchetti2014a}.

The breaking of B1 approximation for the range of temperatures explored in curve $A$ was expected from \cite{meyer1981}\cite{meyer1982}, though, to our knowledge, this is the first time that the effect on the spin relaxation time is calculated and demonstrated to be substantial. We wish now to explore this effect for a different range of temperatures and density, so to properly sample region II of Fig. \ref{fig:region}. We consider another line in region $II$ of Fig.~\ref{fig:region}, denoted by $B$. This corresponds to the parameters: $n_{\mathrm{e}}= 1.5 \times 10^{17} $ $cm^{-3}$, $T_{\mathrm{F}} = 186.6 $ $\mathrm{K}$ and temperature range $T = 70 \div 120$ $\mathrm{K}$. Given the higher Fermi temperature, curve $B$ indeed corresponds to a similar degeneracy regime to curve $A$, and, as previously demonstrated, the Born series  should converges for its range of parameters, see Fig.~\ref{fig:upperBound}. It is then justified for us to compute the SRT in B1 and B2 along curve $B$.

\begin{figure}
\resizebox{0.50\textwidth}{!}{%
  \includegraphics{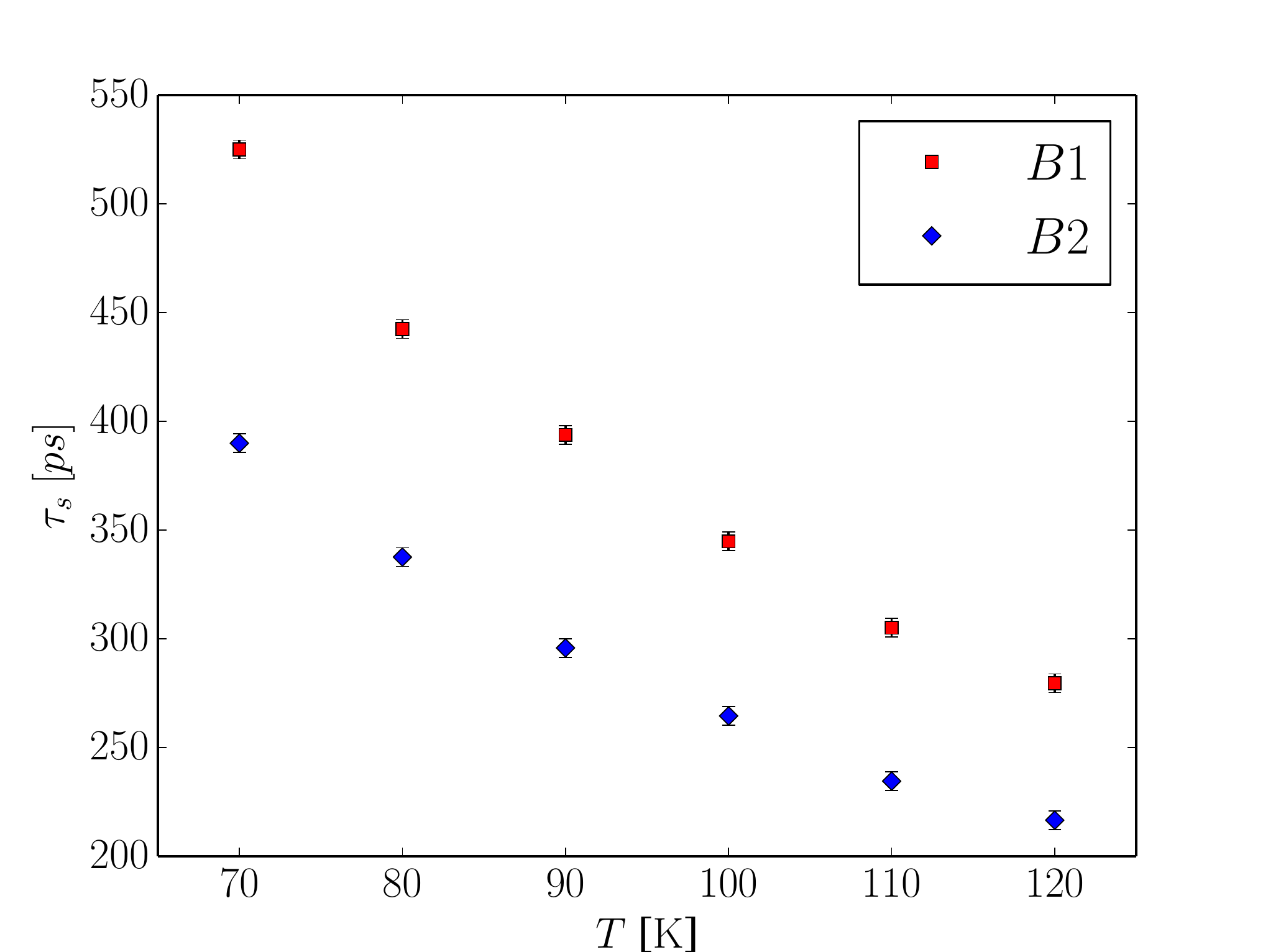}
}
\caption{Comparison of the spin relaxation times $\tau_s$ along line $B$ for $B1$ (squares) and $B2$ (diamonds)
approximations. Here $n_{\mathrm{e}}=
1.5 \times 10^{17}  \mathrm{cm}^{-3}$, $N= 25,000$ and $\gamma_{so}= 21.9 \, \mathrm{eV} \angstrom ^{3}$.}
\label{fig:SRTcurve1}       
\end{figure}

The  results for spin relaxation times $\tau_s$ in B1 and B2 are shown in Fig.~\ref{fig:SRTcurve1}.
Once more we find that the correction to $\tau_s$ due to B2 is quite substantial, with B1 predictions overshooting the B2 results by 35\% at $T = 70$ $\mathrm{K}$ and by 27\% at $T = 120$ $\mathrm{K}$.

As for line $A$, the spin relaxation times $\tau_s^{B1}$  and $\tau_s^{B2}$ along line $B$ decrease monotonically with temperatures, as expected.

For the same temperature, SRT are now more than 20\% greater than the corresponding ones along line $A$, for instance, compare their values for $T=70$ $\mathrm{K}$. An increase of
SRT with the electronic density is a clear signature of being far from the degenerate regime carriers' dynamics \cite{jiang2009}\cite{marchetti2014}.

\section{Conclusion} \label{sec:5}
We have extended ensemble Monte Carlo techniques to include second order Born approximation for the electron-impurity screening and performed a systematic analysis of the effect of using different Born approximations on electronic spin relaxation. 
Our findings demonstrate that, for a quantitative estimate of spin relaxation properties in semiconductors in an intermediate regime of degeneracy ($T\ge 50 $ $\mathrm{K}$, $T/T_F<90\%$), it is crucial to go beyond the first Born approximation. Our results in fact show a substantial difference (more than 30\% for most of the parameter space explored) in spin relaxation times when computed in first and second Born approximations. It is important to recall here that two important requirements for the validity of Stern's FSR formula are met in our simulations: thermal equilibrium and parabolicity. Patterson and Lehoczky approximation may become more important for lower temperatures and should be further investigated including inelastic processes with longitudinal acoustic phonons.
It is worth recalling that we performed EMC calculations of $\tau_s$ for a specific value of SOC ($\gamma_{so}= 21.9 \, \mathrm{eV} \angstrom ^{3}$) which gave us  reliable values, when compared to experiments, for spin relaxation times at room and higher temperatures \cite{marchetti2014}\cite{marchetti2014a}. However the spin-orbit coupling  for GaAs found in the literature varies greatly \cite{fu2008} giving rise to a corresponding large variation for $\tau_s$ values \cite{marchetti2014}\footnote{ For instance, Jiang and Wu used a SOC value of $ 23.9 \, \mathrm{eV} \angstrom ^{3}$ for their electron spin relaxation calculations by microscopic kinetic spin Bloch equation approach \cite{jiang2009}.}: new experimental observations of spin dynamics in doped GaAs in the range of temperatures and doping concentrations carefully chosen in the present work would be useful both to provide a quantitative confirmation to our findings and to the importance of going beyond first-order Born approximation for quantitative estimates of spin properties in the intermediate degeneracy regime.

\begin{acknowledgments}
G. M is grateful to Giovanni Vignale for an enlightening discussion. The authors would also like to thank Matthew Hodgson for his contributions to previous versions of the computational code.
\end{acknowledgments}

\bibliography{prb_references}{}

\begin{thebibliography}{41}%
\makeatletter
\providecommand \@ifxundefined [1]{%
 \@ifx{#1\undefined}
}%
\providecommand \@ifnum [1]{%
 \ifnum #1\expandafter \@firstoftwo
 \else \expandafter \@secondoftwo
 \fi
}%
\providecommand \@ifx [1]{%
 \ifx #1\expandafter \@firstoftwo
 \else \expandafter \@secondoftwo
 \fi
}%
\providecommand \natexlab [1]{#1}%
\providecommand \enquote  [1]{``#1''}%
\providecommand \bibnamefont  [1]{#1}%
\providecommand \bibfnamefont [1]{#1}%
\providecommand \citenamefont [1]{#1}%
\providecommand \href@noop [0]{\@secondoftwo}%
\providecommand \href [0]{\begingroup \@sanitize@url \@href}%
\providecommand \@href[1]{\@@startlink{#1}\@@href}%
\providecommand \@@href[1]{\endgroup#1\@@endlink}%
\providecommand \@sanitize@url [0]{\catcode `\\12\catcode `\$12\catcode
  `\&12\catcode `\#12\catcode `\^12\catcode `\_12\catcode `\%12\relax}%
\providecommand \@@startlink[1]{}%
\providecommand \@@endlink[0]{}%
\providecommand \url  [0]{\begingroup\@sanitize@url \@url }%
\providecommand \@url [1]{\endgroup\@href {#1}{\urlprefix }}%
\providecommand \urlprefix  [0]{URL }%
\providecommand \Eprint [0]{\href }%
\providecommand \doibase [0]{http://dx.doi.org/}%
\providecommand \selectlanguage [0]{\@gobble}%
\providecommand \bibinfo  [0]{\@secondoftwo}%
\providecommand \bibfield  [0]{\@secondoftwo}%
\providecommand \translation [1]{[#1]}%
\providecommand \BibitemOpen [0]{}%
\providecommand \bibitemStop [0]{}%
\providecommand \bibitemNoStop [0]{.\EOS\space}%
\providecommand \EOS [0]{\spacefactor3000\relax}%
\providecommand \BibitemShut  [1]{\csname bibitem#1\endcsname}%
\let\auto@bib@innerbib\@empty
\bibitem [{\citenamefont {Chattopadhyay}\ and\ \citenamefont
  {Queisser}(1981)}]{chattopadhyay1981}%
  \BibitemOpen
  \bibfield  {author} {\bibinfo {author} {\bibfnamefont {D.}~\bibnamefont
  {Chattopadhyay}}\ and\ \bibinfo {author} {\bibfnamefont {H.~J.}\ \bibnamefont
  {Queisser}},\ }\href {\doibase 10.1103/RevModPhys.53.745} {\bibfield
  {journal} {\bibinfo  {journal} {Rev. Mod. Phys.}\ }\textbf {\bibinfo {volume}
  {53}},\ \bibinfo {pages} {745} (\bibinfo {year} {1981})}\BibitemShut
  {NoStop}%
\bibitem [{\citenamefont {Ashcroft}\ and\ \citenamefont
  {Mermin}(1976)}]{ashcroft1976}%
  \BibitemOpen
  \bibfield  {author} {\bibinfo {author} {\bibfnamefont {N.~W.}\ \bibnamefont
  {Ashcroft}}\ and\ \bibinfo {author} {\bibfnamefont {M.~D.}\ \bibnamefont
  {Mermin}},\ }\href@noop {} {\emph {\bibinfo {title} {{Solid State
  Physics}}}}\ (\bibinfo  {publisher} {Saunders College},\ \bibinfo {address}
  {Philadelphia},\ \bibinfo {year} {1976})\BibitemShut {NoStop}%
\bibitem [{\citenamefont {Giuliani}\ and\ \citenamefont
  {Vignale}(2005)}]{giuliani2005}%
  \BibitemOpen
  \bibfield  {author} {\bibinfo {author} {\bibfnamefont {G.}~\bibnamefont
  {Giuliani}}\ and\ \bibinfo {author} {\bibfnamefont {G.}~\bibnamefont
  {Vignale}},\ }\href@noop {} {\emph {\bibinfo {title} {{Quantum Theory of
  Electron Liquid}}}}\ (\bibinfo  {publisher} {Cambridge University Press},\
  \bibinfo {address} {Cambridge},\ \bibinfo {year} {2005})\BibitemShut
  {NoStop}%
\bibitem [{\citenamefont {Sanborn}(1995)}]{sanborn1995}%
  \BibitemOpen
  \bibfield  {author} {\bibinfo {author} {\bibfnamefont {B.~A.}\ \bibnamefont
  {Sanborn}},\ }\href@noop {} {\bibfield  {journal} {\bibinfo  {journal} {Phys.
  Rev. B}\ }\textbf {\bibinfo {volume} {51}},\ \bibinfo {pages} {4601}
  (\bibinfo {year} {1995})}\BibitemShut {NoStop}%
\bibitem [{\citenamefont {Jacoboni}\ and\ \citenamefont
  {Lugli}(1989)}]{jacoboni1989}%
  \BibitemOpen
  \bibfield  {author} {\bibinfo {author} {\bibfnamefont {C.}~\bibnamefont
  {Jacoboni}}\ and\ \bibinfo {author} {\bibfnamefont {P.}~\bibnamefont
  {Lugli}},\ }\href@noop {} {\emph {\bibinfo {title} {{The Monte Carlo Method
  for Semiconductor Device Simulation}}}}\ (\bibinfo  {publisher}
  {Springer-Verlag},\ \bibinfo {address} {Wien New York},\ \bibinfo {year}
  {1989})\BibitemShut {NoStop}%
\bibitem [{\citenamefont {Marchetti}\ \emph
  {et~al.}(2014{\natexlab{a}})\citenamefont {Marchetti}, \citenamefont
  {Hodgson}, \citenamefont {McHugh}, \citenamefont {Chantrell},\ and\
  \citenamefont {D'Amico}}]{marchetti2014}%
  \BibitemOpen
  \bibfield  {author} {\bibinfo {author} {\bibfnamefont {G.}~\bibnamefont
  {Marchetti}}, \bibinfo {author} {\bibfnamefont {M.}~\bibnamefont {Hodgson}},
  \bibinfo {author} {\bibfnamefont {J.}~\bibnamefont {McHugh}}, \bibinfo
  {author} {\bibfnamefont {R.}~\bibnamefont {Chantrell}}, \ and\ \bibinfo
  {author} {\bibfnamefont {I.}~\bibnamefont {D'Amico}},\ }\href@noop {}
  {\bibfield  {journal} {\bibinfo  {journal} {Materials}\ }\textbf {\bibinfo
  {volume} {7}},\ \bibinfo {pages} {2795} (\bibinfo {year}
  {2014}{\natexlab{a}})}\BibitemShut {NoStop}%
\bibitem [{\citenamefont {Marchetti}\ \emph
  {et~al.}(2014{\natexlab{b}})\citenamefont {Marchetti}, \citenamefont
  {Hodgson},\ and\ \citenamefont {D'Amico}}]{marchetti2014a}%
  \BibitemOpen
  \bibfield  {author} {\bibinfo {author} {\bibfnamefont {G.}~\bibnamefont
  {Marchetti}}, \bibinfo {author} {\bibfnamefont {M.}~\bibnamefont {Hodgson}},
  \ and\ \bibinfo {author} {\bibfnamefont {I.}~\bibnamefont {D'Amico}},\
  }\href@noop {} {\bibfield  {journal} {\bibinfo  {journal} {Journal of Applied
  Physics}\ }\textbf {\bibinfo {volume} {16}} (\bibinfo {year}
  {2014}{\natexlab{b}})}\BibitemShut {NoStop}%
\bibitem [{\citenamefont {Omnes}\ and\ \citenamefont
  {Froissart}(1963)}]{omnes1963}%
  \BibitemOpen
  \bibfield  {author} {\bibinfo {author} {\bibfnamefont {R.}~\bibnamefont
  {Omnes}}\ and\ \bibinfo {author} {\bibfnamefont {M.}~\bibnamefont
  {Froissart}},\ }\href@noop {} {\emph {\bibinfo {title} {{Mandelstam Theory
  and Regge Poles}}}}\ (\bibinfo  {publisher} {W.~ A.~ Benjamin, Inc.},\
  \bibinfo {address} {New York},\ \bibinfo {year} {1963})\BibitemShut {NoStop}%
\bibitem [{\citenamefont {Meyer}\ and\ \citenamefont
  {Bartoli}(1982)}]{meyer1982}%
  \BibitemOpen
  \bibfield  {author} {\bibinfo {author} {\bibfnamefont {J.~R.}\ \bibnamefont
  {Meyer}}\ and\ \bibinfo {author} {\bibfnamefont {F.~J.~.}\ \bibnamefont
  {Bartoli}},\ }\href@noop {} {\bibfield  {journal} {\bibinfo  {journal} {Solid
  State Communications}\ }\textbf {\bibinfo {volume} {41}},\ \bibinfo {pages}
  {19 } (\bibinfo {year} {1982})}\BibitemShut {NoStop}%
\bibitem [{\citenamefont {Meyer}\ and\ \citenamefont
  {Bartoli}(1981)}]{meyer1981}%
  \BibitemOpen
  \bibfield  {author} {\bibinfo {author} {\bibfnamefont {J.~R.}\ \bibnamefont
  {Meyer}}\ and\ \bibinfo {author} {\bibfnamefont {F.~J.}\ \bibnamefont
  {Bartoli}},\ }\href@noop {} {\bibfield  {journal} {\bibinfo  {journal} {Phys.
  Rev. B}\ }\textbf {\bibinfo {volume} {24}},\ \bibinfo {pages} {2089}
  (\bibinfo {year} {1981})}\BibitemShut {NoStop}%
\bibitem [{\citenamefont {Joachain}(1987)}]{joachain1987}%
  \BibitemOpen
  \bibfield  {author} {\bibinfo {author} {\bibfnamefont {C.~J.}\ \bibnamefont
  {Joachain}},\ }\href@noop {} {\emph {\bibinfo {title} {{Quantum Collision
  Theory}}}}\ (\bibinfo  {publisher} {North-Holland Physics Publishing},\
  \bibinfo {address} {Amsterdam},\ \bibinfo {year} {1987})\BibitemShut
  {NoStop}%
\bibitem [{\citenamefont {Friedel}(1958)}]{friedel1958}%
  \BibitemOpen
  \bibfield  {author} {\bibinfo {author} {\bibfnamefont {J.}~\bibnamefont
  {Friedel}},\ }\href {\doibase 10.1007/BF02751483} {\bibfield  {journal}
  {\bibinfo  {journal} {Il Nuovo Cimento}\ }\textbf {\bibinfo {volume} {7}},\
  \bibinfo {pages} {287} (\bibinfo {year} {1958})}\BibitemShut {NoStop}%
\bibitem [{\citenamefont {Stern}(1967)}]{stern1967}%
  \BibitemOpen
  \bibfield  {author} {\bibinfo {author} {\bibfnamefont {F.}~\bibnamefont
  {Stern}},\ }\href {\doibase 10.1103/PhysRev.158.697} {\bibfield  {journal}
  {\bibinfo  {journal} {Phys. Rev.}\ }\textbf {\bibinfo {volume} {158}},\
  \bibinfo {pages} {697} (\bibinfo {year} {1967})}\BibitemShut {NoStop}%
\bibitem [{\citenamefont {Patterson}\ and\ \citenamefont
  {Lehoczky}(1989)}]{patterson1989}%
  \BibitemOpen
  \bibfield  {author} {\bibinfo {author} {\bibfnamefont {J.~D.}\ \bibnamefont
  {Patterson}}\ and\ \bibinfo {author} {\bibfnamefont {S.~L.~.}\ \bibnamefont
  {Lehoczky}},\ }\href@noop {} {\bibfield  {journal} {\bibinfo  {journal}
  {Physics Letters A}\ }\textbf {\bibinfo {volume} {137}},\ \bibinfo {pages}
  {137 } (\bibinfo {year} {1989})}\BibitemShut {NoStop}%
\bibitem [{\citenamefont {Zutic}\ \emph {et~al.}(2004)\citenamefont {Zutic},
  \citenamefont {Fabian},\ and\ \citenamefont {Sarma}}]{zutic2004}%
  \BibitemOpen
  \bibfield  {author} {\bibinfo {author} {\bibfnamefont {I.}~\bibnamefont
  {Zutic}}, \bibinfo {author} {\bibfnamefont {J.}~\bibnamefont {Fabian}}, \
  and\ \bibinfo {author} {\bibfnamefont {S.~D.}\ \bibnamefont {Sarma}},\
  }\href@noop {} {\bibfield  {journal} {\bibinfo  {journal} {Rev. Mod. Phys.}\
  }\textbf {\bibinfo {volume} {76}},\ \bibinfo {pages} {323} (\bibinfo {year}
  {2004})}\BibitemShut {NoStop}%
\bibitem [{\citenamefont {Dyakonov}\ and\ \citenamefont
  {Perel}(1971)}]{dyakonov1971}%
  \BibitemOpen
  \bibfield  {author} {\bibinfo {author} {\bibfnamefont {M.~I.}\ \bibnamefont
  {Dyakonov}}\ and\ \bibinfo {author} {\bibfnamefont {V.~I.}\ \bibnamefont
  {Perel}},\ }\href@noop {} {\bibfield  {journal} {\bibinfo  {journal} {Sov.
  Phys. Solid State}\ }\textbf {\bibinfo {volume} {13}},\ \bibinfo {pages}
  {3023} (\bibinfo {year} {1971})}\BibitemShut {NoStop}%
\bibitem [{\citenamefont {Loss}\ and\ \citenamefont
  {DiVincenzo}(1998)}]{loss1998}%
  \BibitemOpen
  \bibfield  {author} {\bibinfo {author} {\bibfnamefont {D.}~\bibnamefont
  {Loss}}\ and\ \bibinfo {author} {\bibfnamefont {D.~P.}\ \bibnamefont
  {DiVincenzo}},\ }\href@noop {} {\bibfield  {journal} {\bibinfo  {journal}
  {Phys. Rev. A}\ }\textbf {\bibinfo {volume} {57}},\ \bibinfo {pages} {120}
  (\bibinfo {year} {1998})}\BibitemShut {NoStop}%
\bibitem [{\citenamefont {Datta}\ and\ \citenamefont {Das}(1990)}]{datta1990}%
  \BibitemOpen
  \bibfield  {author} {\bibinfo {author} {\bibfnamefont {S.}~\bibnamefont
  {Datta}}\ and\ \bibinfo {author} {\bibfnamefont {B.}~\bibnamefont {Das}},\
  }\href@noop {} {\bibfield  {journal} {\bibinfo  {journal} {Applied Physics
  Letters}\ }\textbf {\bibinfo {volume} {56}},\ \bibinfo {pages} {665}
  (\bibinfo {year} {1990})}\BibitemShut {NoStop}%
\bibitem [{\citenamefont {Tsymbal}\ \emph {et~al.}(2003)\citenamefont
  {Tsymbal}, \citenamefont {Mryasov},\ and\ \citenamefont
  {LeClair}}]{tsymbal2003}%
  \BibitemOpen
  \bibfield  {author} {\bibinfo {author} {\bibfnamefont {E.~Y.}\ \bibnamefont
  {Tsymbal}}, \bibinfo {author} {\bibfnamefont {O.~N.}\ \bibnamefont
  {Mryasov}}, \ and\ \bibinfo {author} {\bibfnamefont {P.~R.}\ \bibnamefont
  {LeClair}},\ }\href@noop {} {\bibfield  {journal} {\bibinfo  {journal}
  {Journal of Physics: Condensed Matter}\ }\textbf {\bibinfo {volume} {15}},\
  \bibinfo {pages} {R109} (\bibinfo {year} {2003})}\BibitemShut {NoStop}%
\bibitem [{\citenamefont {Jungwirth}\ \emph {et~al.}(2012)\citenamefont
  {Jungwirth}, \citenamefont {Wunderlich},\ and\ \citenamefont
  {Olejnik}}]{jungwirth2012}%
  \BibitemOpen
  \bibfield  {author} {\bibinfo {author} {\bibfnamefont {T.}~\bibnamefont
  {Jungwirth}}, \bibinfo {author} {\bibfnamefont {J.}~\bibnamefont
  {Wunderlich}}, \ and\ \bibinfo {author} {\bibfnamefont {K.}~\bibnamefont
  {Olejnik}},\ }\href@noop {} {\bibfield  {journal} {\bibinfo  {journal}
  {Nature Materials}\ }\textbf {\bibinfo {volume} {11}},\ \bibinfo {pages}
  {382} (\bibinfo {year} {2012})}\BibitemShut {NoStop}%
\bibitem [{\citenamefont {Kikkawa}\ and\ \citenamefont
  {Awschalom}(1998)}]{kikkawa1998}%
  \BibitemOpen
  \bibfield  {author} {\bibinfo {author} {\bibfnamefont {J.~M.}\ \bibnamefont
  {Kikkawa}}\ and\ \bibinfo {author} {\bibfnamefont {D.~D.}\ \bibnamefont
  {Awschalom}},\ }\href@noop {} {\bibfield  {journal} {\bibinfo  {journal}
  {Phys. Rev. Lett.}\ }\textbf {\bibinfo {volume} {80}},\ \bibinfo {pages}
  {4313} (\bibinfo {year} {1998})}\BibitemShut {NoStop}%
\bibitem [{\citenamefont {Langer}\ and\ \citenamefont
  {V.~Ambegaokar}(1961)}]{langer1961}%
  \BibitemOpen
  \bibfield  {author} {\bibinfo {author} {\bibfnamefont {J.~. S.~.}\
  \bibnamefont {Langer}}\ and\ \bibinfo {author} {\bibfnamefont
  {V.}~\bibnamefont {V.~Ambegaokar}},\ }\href@noop {} {\bibfield  {journal}
  {\bibinfo  {journal} {Phys. Rev.}\ }\textbf {\bibinfo {volume} {121}},\
  \bibinfo {pages} {1090} (\bibinfo {year} {1961})}\BibitemShut {NoStop}%
\bibitem [{\citenamefont {Mahan}(2000)}]{mahan2000}%
  \BibitemOpen
  \bibfield  {author} {\bibinfo {author} {\bibfnamefont {G.~D.}\ \bibnamefont
  {Mahan}},\ }\href@noop {} {\emph {\bibinfo {title} {{Many-Particle
  Physics}}}}\ (\bibinfo  {publisher} {Kluwer Academic},\ \bibinfo {address}
  {New York},\ \bibinfo {year} {2000})\BibitemShut {NoStop}%
\bibitem [{Note1()}]{Note1}%
  \BibitemOpen
  \bibinfo {note} {In the degenerate case ($T=0$ $\protect \mathrm {K}$) the
  phase shifts must be evaluated at Fermi energy $E_\protect \mathrm {F}$ or
  equivalently at Fermi wavector $k_\protect \mathrm {F}$.}\BibitemShut {Stop}%
\bibitem [{\citenamefont {Vurgaftman}\ \emph {et~al.}(2001)\citenamefont
  {Vurgaftman}, \citenamefont {Meyer},\ and\ \citenamefont
  {Ram-Mohan}}]{vurgaftman2001}%
  \BibitemOpen
  \bibfield  {author} {\bibinfo {author} {\bibfnamefont {I.}~\bibnamefont
  {Vurgaftman}}, \bibinfo {author} {\bibfnamefont {J.~R.}\ \bibnamefont
  {Meyer}}, \ and\ \bibinfo {author} {\bibfnamefont {L.~R.}\ \bibnamefont
  {Ram-Mohan}},\ }\href@noop {} {\bibfield  {journal} {\bibinfo  {journal}
  {Journal of Applied Physics}\ }\textbf {\bibinfo {volume} {89}},\ \bibinfo
  {pages} {5815} (\bibinfo {year} {2001})}\BibitemShut {NoStop}%
\bibitem [{\citenamefont {Nersisyan}\ and\ \citenamefont
  {Fernandez-Varea"}(2013)}]{nersisyan2013}%
  \BibitemOpen
  \bibfield  {author} {\bibinfo {author} {\bibfnamefont {H.~B.}\ \bibnamefont
  {Nersisyan}}\ and\ \bibinfo {author} {\bibfnamefont {J.~M.}\ \bibnamefont
  {Fernandez-Varea"}},\ }\href@noop {} {\bibfield  {journal} {\bibinfo
  {journal} {Nuclear Instruments and Methods in Physics Research Section B:
  Beam Interactions with Materials and Atoms}\ }\textbf {\bibinfo {volume}
  {311}},\ \bibinfo {pages} {121 } (\bibinfo {year} {2013})}\BibitemShut
  {NoStop}%
\bibitem [{Note2()}]{Note2}%
  \BibitemOpen
  \bibinfo {note} {It is worthwhile to recall here that in general if the true
  and the first Born approximation phase shifts are small this indeed does not
  imply the validity of the Born approximation for a general short-range
  potential \cite {peierls1979}.}\BibitemShut {Stop}%
\bibitem [{\citenamefont {Blakemore}(1962)}]{blakemore1962}%
  \BibitemOpen
  \bibfield  {author} {\bibinfo {author} {\bibfnamefont {J.~S.}\ \bibnamefont
  {Blakemore}},\ }\href@noop {} {\emph {\bibinfo {title} {{Semiconductor
  Statistics}}}}\ (\bibinfo  {publisher} {Pergamon},\ \bibinfo {year}
  {1962})\BibitemShut {NoStop}%
\bibitem [{\citenamefont {Dingle}(1955)}]{dingle1955}%
  \BibitemOpen
  \bibfield  {author} {\bibinfo {author} {\bibfnamefont {R.~B.}\ \bibnamefont
  {Dingle}},\ }\href@noop {} {\bibfield  {journal} {\bibinfo  {journal}
  {Philos. Mag.}\ }\textbf {\bibinfo {volume} {46}},\ \bibinfo {pages} {4601}
  (\bibinfo {year} {1955})}\BibitemShut {NoStop}%
\bibitem [{\citenamefont {Jiang}\ and\ \citenamefont {Wu}(2009)}]{jiang2009}%
  \BibitemOpen
  \bibfield  {author} {\bibinfo {author} {\bibfnamefont {J.~H.}\ \bibnamefont
  {Jiang}}\ and\ \bibinfo {author} {\bibfnamefont {M.~W.}\ \bibnamefont {Wu}},\
  }\href@noop {} {\bibfield  {journal} {\bibinfo  {journal} {Phys. Rev. B}\
  }\textbf {\bibinfo {volume} {79}},\ \bibinfo {pages} {125206} (\bibinfo
  {year} {2009})}\BibitemShut {NoStop}%
\bibitem [{\citenamefont {Fabian}\ \emph {et~al.}(2007)\citenamefont {Fabian},
  \citenamefont {Matos-Abiague}, \citenamefont {Ertler}, \citenamefont
  {Stano},\ and\ \citenamefont {Zutic}}]{fabian2007}%
  \BibitemOpen
  \bibfield  {author} {\bibinfo {author} {\bibfnamefont {J.}~\bibnamefont
  {Fabian}}, \bibinfo {author} {\bibfnamefont {A.}~\bibnamefont
  {Matos-Abiague}}, \bibinfo {author} {\bibfnamefont {C.}~\bibnamefont
  {Ertler}}, \bibinfo {author} {\bibfnamefont {P.}~\bibnamefont {Stano}}, \
  and\ \bibinfo {author} {\bibfnamefont {I.}~\bibnamefont {Zutic}},\
  }\href@noop {} {\bibfield  {journal} {\bibinfo  {journal} {Acta Phys. Slov.}\
  }\textbf {\bibinfo {volume} {57}},\ \bibinfo {pages} {565} (\bibinfo {year}
  {2007})}\BibitemShut {NoStop}%
\bibitem [{Note3()}]{Note3}%
  \BibitemOpen
  \bibinfo {note} {In a semiclassical picture the substantial contribution to
  e-i scattering comes from waves for which $l\leq kb$ where $b$ is the impact
  parameter and $k$ the electron wavevector. \cite {joachain1987}. For
  instance, assuming that $b$ is half the interatomic distance between the
  impurities and $k\simeq k_F$, the Fermi momentum, one can see that only few
  waves in general matter for slow electrons.}\BibitemShut {Stop}%
\bibitem [{Note4()}]{Note4}%
  \BibitemOpen
  \bibinfo {note} {To date, Brooks-Herring approach is one of most reliable and
  successful physical picture of electron-ion interaction. The other widely
  used model, by Conwell and Weisskopf \cite {chattopadhyay1981}, is instead
  based upon the unscreened Coulomb which is inconsistent with
  RPA.}\BibitemShut {Stop}%
\bibitem [{Note5()}]{Note5}%
  \BibitemOpen
  \bibinfo {note} {Note that from the kinematics of non-relativistic collisions
  one can show that the scattering angle $\theta $ in laboratory frame of
  reference is very close to the corresponding scattering angle $\theta _{CoM}$
  in centre of mass frame of reference, i.e. $\theta \approx \theta _{CoM}$
  because $m^{\ast }\ll M_I$, being $M_I$ the mass of impurity centers \cite
  {kamal2014} Moreover the maximum scattering angle $\theta _{max}$ is obtained
  when $\theta =\pi $ \cite {kamal2014}.}\BibitemShut {Stop}%
\bibitem [{Note6()}]{Note6}%
  \BibitemOpen
  \bibinfo {note} {The EMC algorithm requires that the carriers also undergo to
  a self-scattering where nothing actually happens \cite
  {jacoboni1989}.}\BibitemShut {Stop}%
\bibitem [{\citenamefont {Calogero}(1965)}]{calogero1965}%
  \BibitemOpen
  \bibfield  {author} {\bibinfo {author} {\bibfnamefont {F.}~\bibnamefont
  {Calogero}},\ }\href@noop {} {\bibfield  {journal} {\bibinfo  {journal}
  {Comm. Math. Phys.}\ }\textbf {\bibinfo {volume} {1}},\ \bibinfo {pages} {80}
  (\bibinfo {year} {1965})}\BibitemShut {NoStop}%
\bibitem [{\citenamefont {Kohn}(1954)}]{kohn1954}%
  \BibitemOpen
  \bibfield  {author} {\bibinfo {author} {\bibfnamefont {W.}~\bibnamefont
  {Kohn}},\ }\href@noop {} {\bibfield  {journal} {\bibinfo  {journal} {Rev.
  Mod. Phys.}\ }\textbf {\bibinfo {volume} {26}},\ \bibinfo {pages} {292}
  (\bibinfo {year} {1954})}\BibitemShut {NoStop}%
\bibitem [{\citenamefont {Fu}\ \emph {et~al.}(2008)\citenamefont {Fu},
  \citenamefont {Weng},\ and\ \citenamefont {Wu}}]{fu2008}%
  \BibitemOpen
  \bibfield  {author} {\bibinfo {author} {\bibfnamefont {J.~Y.}\ \bibnamefont
  {Fu}}, \bibinfo {author} {\bibfnamefont {M.~Q.}\ \bibnamefont {Weng}}, \ and\
  \bibinfo {author} {\bibfnamefont {M.~W.}\ \bibnamefont {Wu}},\ }\href@noop {}
  {\bibfield  {journal} {\bibinfo  {journal} {Physica E: Low-dimensional
  Systems and Nanostructures}\ }\textbf {\bibinfo {volume} {40}},\ \bibinfo
  {pages} {2890 } (\bibinfo {year} {2008})}\BibitemShut {NoStop}%
\bibitem [{Note7()}]{Note7}%
  \BibitemOpen
  \bibinfo {note} {For instance, Jiang and Wu used a SOC value of $ 23.9
  \protect \tmspace +\thinmuskip {.1667em} \protect \mathrm {eV} \unhbox
  \voidb@x \hbox {\protect \normalfont \r A}^{3}$ for their electron spin
  relaxation calculations by microscopic kinetic spin Bloch equation approach
  \cite {jiang2009}.}\BibitemShut {Stop}%
\bibitem [{\citenamefont {Peierls}(1979)}]{peierls1979}%
  \BibitemOpen
  \bibfield  {author} {\bibinfo {author} {\bibfnamefont {R.}~\bibnamefont
  {Peierls}},\ }\href@noop {} {\emph {\bibinfo {title} {{Surprises in
  Theoretical Physics}}}}\ (\bibinfo  {publisher} {Princeton University
  Press},\ \bibinfo {year} {1979})\BibitemShut {NoStop}%
\bibitem [{\citenamefont {Kamal}(2014)}]{kamal2014}%
  \BibitemOpen
  \bibfield  {author} {\bibinfo {author} {\bibfnamefont {A.}~\bibnamefont
  {Kamal}},\ }\href@noop {} {\emph {\bibinfo {title} {{Nuclear Physics}}}}\
  (\bibinfo  {publisher} {Springer-Verlag},\ \bibinfo {address} {Berlin
  Heidelberg},\ \bibinfo {year} {2014})\BibitemShut {NoStop}%
\end{thebibliography}%

\bibliographystyle{apsrev4-1}

\end{document}